# Drone Control based on Mental Commands and Facial Expressions

Iuliana MARIN
*Faculty of Engineering in Foreign Languages*
*University POLITEHNICA of Bucharest*
Bucharest, Romania
marin.iulliana25@gmail.com

Myssar Jabbar Hammood AL-BATTBOOTTI
*Faculty of Engineering in Foreign Languages*
*University POLITEHNICA of Bucharest*
Bucharest, Romania
jmyssar@gmail.com

Nicolae GOGA
*Faculty of Engineering in Foreign Languages*
*University POLITEHNICA of Bucharest*
Bucharest, Romania
n.goga@rug.nl

*Abstract*—When it is tried to control drones, there are many different ways through various devices, using either motions like facial motion, special gloves with sensors, red, green, blue cameras on the laptop or even using smartwatches by performing gestures that are picked up by motion sensors. The paper proposes a work on how drones could be controlled using brainwaves without any of those devices. The drone control system of the current research was developed using electroencephalogram signals took by an Emotiv Insight headset. The electroencephalogram signals are collected from the user's brain. The processed signal is then sent to the computer via Bluetooth. The headset employs Bluetooth Low Energy for wireless transmission. The brain of the user is trained in order to use the generated electroencephalogram data. The final signal is transmitted to Raspberry Pi zero via the MQTT messaging protocol. The Raspberry Pi controls the movement of the drone through the incoming signal from the headset. After years, brain control can replace many normal input sources like keyboards, touch screens or other traditional ways, so it enhances interactive experiences and provides new ways for disabled people to engage with their surroundings.

*Keywords—Headset; electroencephalogram; drone; mental commands; facial expressions.*

## I. Introduction

As the world evolves, mind control technology through the interactive system between the brain with any physical machine, allows the human beings to look forward to acquiring new devices with new technology to help them move around, control things via thinking like a wheelchair, robotic arm, robots in a comfortable manner.

The Brain-Computer Interface (BCI) is one of the world's most powerful and advanced technologies that will be the solution for many people around the world in many different fields. This includes an integration of the brain and machine, both sharing an interface to make the communication channel among the brain and an item to be controlled externally. The human brain has a neural network which contains groups of neurons that are connected to each other and they are responsible for the transmission of brain activity. Hans Berger, a German psychiatrist, was the first person who measured the brain waves in 1924, using the electroencephalography, a method for recording the brain waves. This invention is now known as the EEG device which uses the brain waves signals produced by the neurons, transmits them to a computer which interprets the signals into data [1]. These data are translated to control a device which is connected to the computer. Electrodes are used to measure the brain electrical activity by using the Emotiv Insight headset. There are several hesitancy packs composing the EEG signals, namely Delta δ, Beta β, Theta θ, Alpha α, and Gamma γ. Each pack reflects a different brain activity [2]. The power of these bands is regularly changed during the same day. In fact, the strength of different EEG bands is correlated with the brain activity and its degree of awareness.

This research deals with the development of a BCI system between the Parrot Mambo MiniDrone and the human brainwave to control the drone's movements. BCI allows the direct connection between the drone and the brain activity. The brain activity is analyzed using electroencephalogram signals, which are also referred to as brainwaves. The Emotiv Insight headset is the tool used to capture the EEG signal. The headset can wirelessly convey the EEG signal to the laptop via Bluetooth. The EEG signals are processed and transformed into mental controls. According to the acquired mental command (e.g. up, down, left, right), the electrical output signal is sent to the drone to execute the required movement. A computer program is used to translate the EEG signal into mental control and the wireless transmission of the command is sent towards the Parrot Mambo MiniDrone. In the next section are presented the literature review regarding previous work on EEG, the proposed system of drone control using human brainwave, along with further information about the used components, namely the EEG headset, Raspberry Pi, BCI system. Section 3 outlines the results after testing the system. The last section outlines the conclusions and future work.

## II. Methodology

Day by day the world depends on drones in many fields and they become more popular in this decade. The use of drones is useful and helpful in many fields, such as the search and rescue operations, delivering goods, mail and food, forest fires, emergency medicine delivery. In these fields the drones should be controlled by a device. There are many ways of controlling the drones, but the current research focuses on the most creative one, namely the mind control via brainwaves.

The current research aims to support the brain usage to analyze digital information and get the outcomes, letting the user visualize and hear without using the eyes and ears, such that deaf and blind people can regain their lost sensations. BCI can assist the persons who are immovable to switch gadgets over and to interconnect with people. There are persons who agonized numerous sicknesses and they caused them to become immovable, partaking in the progress of innovative BCI systems.

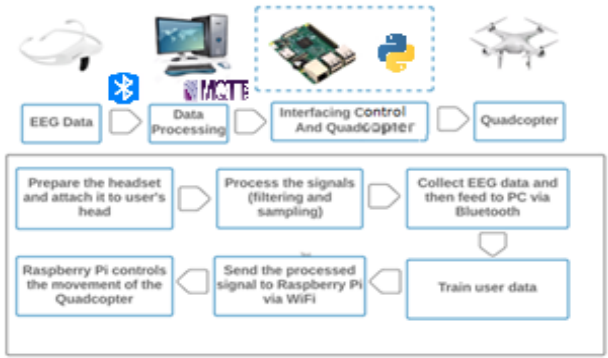

Fig. 1. *BCI system*

Nowadays, the Internet of Things applications start to be widespread and more attention needs to be paid in this direction for research purposes in order to study brainwaves, as well as to interconnect devices which can be bought from market.

*A. Previous work on EEG*

Om Raheja studied different kinds of drone systems and used a flight controller, with preinstalled craft types and an automatic calibration function [3]. A zigbee module was used in order to achieve a wireless link between the sensor used by the user and the quadrotor drone. Zigbee was connected to the Arduino nano controller present on the receiving end.

Pan Peining used other kind of EEG headset and compared it with Emotiv Insight headset [4]. The focus was put on the motion sensors, battery power, facial expressions and the supported platforms. Sim Kok Swee developed of a brainwaves wheelchair [5]. The primary goal was to build a wheelchair that was brain-controlled without physical feedback to control the user's input. In this research they used the Emotiv Epoc headset with Arduino Uno programmed with C.

Compared to previous presented solutions, Raspberry Pi can be programmed using the Python programming language. The greatest benefit Raspberry Pi has over Arduino and Zigbee is its much higher computing power, it is more complex and algorithms can be run better on Raspberry Pi.

*B. The proposed system*

This research developed a BCI system to control a drone. The Emotiv Insight EEG headset used the brainwaves indicators pattern which were involved while controlling the drone. The EEG headset transmits the commands to Raspberry Pi. A connection channel is created through wireless transmission via Bluetooth. The drone is controlled by processing using Raspberry Pi.

Drone control using human brainwave is useful for connecting devices between them, such that without any physical human actions, the drone can be moved in the desired position. Fig. 1 displays the BCI system . The data from the EEG are gathered using the Emotiv Insight headset and transferred via Bluetooth to a computer for data processing using the Python programming language.

The dedicated computer program implements information handling algorithms to analyze EEG information in real time. The outcomes of the processing are allocated to the control commands of the drone and sent via the MQTT messaging protocol to the Raspberry Pi board, which directs the outputs of the digital computer to analog voltage signals that can be detected by the drone. This process is done using Python. The following subsections describe the components of the system.

*C. EEG headset*

The Emotiv Insight headset has five wireless channels which are distributed around the brain, covering the front, back and the sides of the head, as shown in Fig. 2. All these channels record and interpret waves from the brain into significant information that can be comprehended.

The headset has sophisticated electronics designed for daily use that are fully optimized to generate smooth, robust signals whenever and wherever Emotiv Insight measures and tracks focus, engagement, interest, excitement, relaxation, and stress levels [6]. Moreover, Emotiv Insight interprets fundamental mental commands, such as pushing, pulling, levitating, rotating. It also detects facial expressions, such as blink, wink, frown, surprise, and smile.

Emotiv Insight has a number of characteristics. It consists of five EEG information channels, more than other commercial systems. It includes two electrodes of reference for the cancellation of noise. Each channel has a 128 Hz data transfer rate. It has dry electrodes that do not need to be moist with saline alternatives in further preparing.

A Bluetooth connection enables information to be received directly on a cable-free laptop. It utilizes an inner lithium battery with a duration of at least 4 hours. Three inertial sensors are included, namely accelerometer, gyroscope, and magnetometer.

*D. Raspberry Pi*

Raspberry Pi is a one-board computer that is commonly used for education. Raspberry Pi usually has an operating system based on Linux. Overall Raspberry is asserted to perform any job that a common desktop PC can perform.

Raspberry Pi Zero was selected to be used for testing due to the small dimensions and can be placed on the drone. It has HDMI out, 1 GHz processor and 512 MG RAM, being functional and at a low cost.

*E. BCI System*

The BCI system gathers signals from the brainwaves. It analyzes and interprets them into instructions that are delivered to a physical device to perform the desired action after acquiring the signals. Therefore, BCIs do not use ordinary nerve and muscle output pathways of the brain.

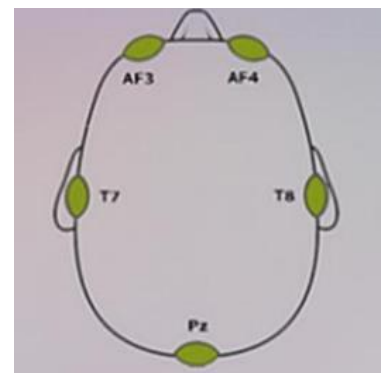

Fig. 2. *Emotiv Insight headset terminals locations*

The BCI system three parts: signal or data acquisition, signal processing (function extraction, translation of features), and output device. These parts are governed by a protocol that specifies operation timing, signal handling information, device command nature and efficiency. For the current research, this is done using a commercially accessible EEG-based BCI device named Emotiv Insight was used. It offers 5-channels using 5 electrodes to collect signals from the head surface that are adequate for this construction.

*Emotiv Insight, EEG reader device*

Emotiv Insight connects to the desktop via Bluetooth and the EEG signals can be read from the brain of the human. EEG wave channels read from the user input. The user needs to get trained in order to match or map EEG information to recognize commands like 'move backwards',' move forward', 'turn left', 'turn right'. For example, as the user thinks about accelerating the drone when training the Insight to learn the 'move forward' command, it corresponds to the EEG signal commands which already is known by the headset. Once the training for the mental commands is completed, when the user thinks about them, the device knows to recognize them.

*Brain Computer Interface Layout*

Signal acquisition in a BCI system uses a sensor modality to help the brain signals measurements, by using multi-electrode arrays. The signals are digitized and transferred to a computer after being amplified and filtered.

BCI feature extraction evaluates, represents and analyzes digital signals to become output instructions. This functionality is correlated with the intention of the user. The resulting signal characteristics are the inputs of the translation function algorithm, which transforms the characteristics to device instructions, like in the case of a cursor control, letter selection, robotic arm operation. It then provides the client with feedback, completing the closed loop of BCI.

*EEG based non-invasive BCI*

EEG uses a conductive gel or paste for recording through the electrodes positioned on the scalp. The number of electrodes used by the present EEG BCI ranges from just a few to 100 electrodes. Because electrode gel can dry out and the setup operation must be repeated before each BCI use session, this EEG-based BCI is not appropriate. Dry electrode arrays may be used as a feasible alternative.

To get better recordings, by following the International 10-20 system, the electrodes need to be positioned correctly, using the lobes, namely 'F' for frontal lobe, 'P' stands for parietal lobe, 'C' stands for central sulcus and 'O' stands for occipital lobe.

First, the user wears the headset, and ensures the locations of the sensors using the head-set software, as well as good signal quality at the same time. The signal is sampled at 128 or 256 samples per second per channel and the bandpass is filtered between 0.2 Hz and 43 Hz [7]. The EEG information obtained from the headset is transferred via a 2.4 G Hz Wi-Fi channel to a data processing device, namely the associated computer. The received data are processed to remove noise. The Raspberry Pi status is checked to find if it is on or off.

If Raspberry Pi is on, the signal is transformed into a message to Raspberry Pi. When the message reaches destination, the content is verified. Each message belongs to a certain condition. If the Raspberry Pi received "Fw" message, the motors are turned on and the drone moves forward. In addition, if the Raspberry Pi receives "Bw" the message, the motors are turned on and then the drone moves backward. If Raspberry Pi receives the "Right" message, the left motor is turned on and the drone turns right and if it receives "Left" message, the drone turns left. If the Raspberry Pi received a "stop" message, the drone will land.

III. RESULTS

Emotiv allows the user to manage mental commands. Every user profile can create unique commands. Each command has a label (for instance, left, right) and a connection to a custom liveliness that can be performed. Emotiv Insight can push, pull, lift, drop, left, right, rotate left, rotate right, rotate forwards, rotate backwards, rotate clockwise, rotate anticlockwise the drone. All these commands are coded.

The initial phase in making mental orders is to prepare the framework to perceive the essential mental command, the supposed neutral condition, by recording for a while the user's cerebrum designs, while not thinking of any order.

Preparing another psychological order is as straightforward as choosing the ideal order name in preparing mode and afterward envisioning the outcomes of the order for 8 seconds (for instance, envision the object and perform the command) while the framework is recording the psychological examples you need to connect with the order.

The request is then live and it can be tested and practiced. After a couple of rehashed attempts and the same number of preparing refreshes as it is needed, the request is prepared to be utilized and the synopsis of preparing information is put away in the client profile.

The more the user gets trained, the better the framework will have the option to distinguish the example of cerebrum action related with the user's thought and the better she/he will figure out how to reproduce that idea in their psyche. At the point when a dynamic order is initiated by the user, the wanted result or movement will be triggered.

The communication between the Parrot Mambo Fly drone and Raspberry Pi Zero is done by using the MQTT protocol which is a machine-to-machine connectivity protocol. This is based on publish/subscribe message transportation. This protocol is used for Internet of Things , as well as it is utilized by healthcare provides, for home automation, providing a good distribution of information to one or multiple receivers.

The software application was developed using Python version 3 and the Emotiv Cortex that allows the connection between the computer and the headset. When the headset is charged and connected, the test can take place. Web sockets and JSON-RPC (remote procedure call protocol in JSON) technologies are used for linking the computer and the Emotiv Insight headset.

Web sockets store data for the program and the headset to send and receive input and output. JSON-RPC is responsible of data transmission. The Emotiv Cortex is downloaded from https://www.emotiv.com/developer/ and then it is installed on the computer. The software is available for Windows and macOS.

The Cortex application needs to be registered from the personal account, at the Cortex Apps section. In this way, the connectivity of the headset and the headset is done based on the provided credentials, namely the application name, the client id and the client secret code. The connection to the Emotiv headset starts by getting authorized based on the user login credentials, after which the information about the Cortex is received. When the JSON data is received from the headset, the "fac" corresponding to facial expression and "com" corresponding to mental command are both analyzed.

The mental commands can be neutral when everything remains as it is (as in Fig. 3), push in order to move object forward, lift for making the object go up, drop to make the object go down and left to move the object to the left side. The output coming from the headset regarding the facial expressions and the mental commands is displayed in Fig. 3.

The flying drone while flying upwards along with the five channels of EEG waves can be visualized in Fig. 4. The AF3 electrode monitors the activity of the left frontal region, T7 and T8 analyze the temporal lobe, Pz corresponds to the electrode placed on the center of the user's scalp to ensure a better fitting on hair [8]. The AF4 electrode observes the right frontal scalp activity. The same is done for the facial expressions which can be neutral, smile, frown, clench, surprise, left or right wink and blink.

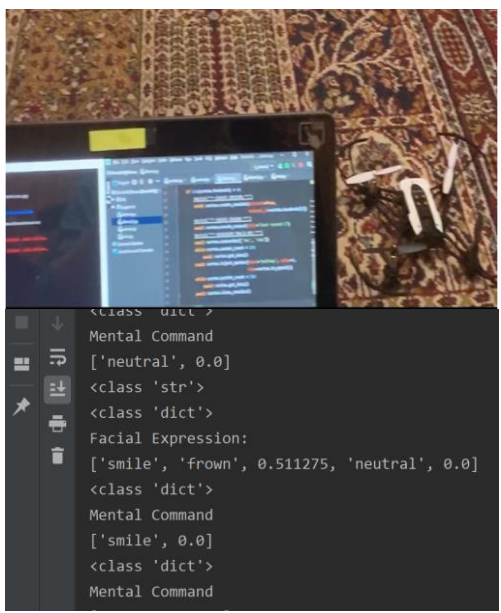

Fig. 3.  *Output of the triggered facial expressions and mental commands*

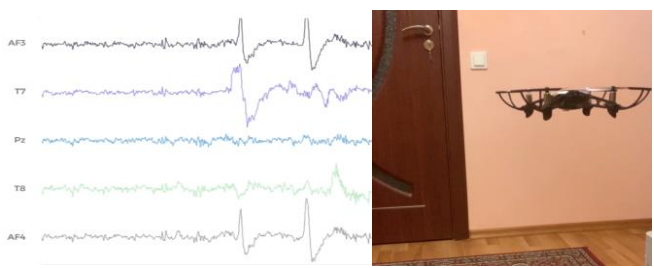

Fig. 4.  *Drone flying up while being controlled by headset commands, along with the five channels of the EEG waves*

The mental command argument value is analyzed and a message is published having as parameter the provided command. During the test of the application, the output was in 88% of the cases right. This was caused by the fact that the headset user needs to really be concentrated and trained while testing the software application. If the user is untrained, the accuracy percentage will be significantly lower.

IV. CONCLUSIONS

In conclusion the Emotiv Insight headset was successfully tested for getting information about the facial expressions and the mental commands. The results match better the reality when the user trains herself/himself by using the EmotivBCI software. The program received information from the Emotiv Insight headset just after the Emotiv Cortex software application was successfully installed on the computer. The software runs on Windows and macOS. The user needs to have a personal account for the Cortex application.

The connection to the headset is done just after the credentials are recognized, namely the application name, the client id and the client secret code. While using the headset and the software application, the user needs to be really concentrated. Otherwise, the precision will decrease. During the tests which were done, the precision was of 88% based on the accuracy of the mental commands. The results depend very much on the previous training which is done for the mental commands which are wanted to be interpreted.

As future work, it would be useful to use the mental commands and control the movements of the drone. The next improvements of the system consist in creating the connection to the drone and to send requests to it based on the mental commands.